\documentclass[aps,prl,reprint,groupedaddress]{revtex4-2}
\usepackage{graphicx}
\usepackage{amsmath}
\usepackage[euler]{textgreek}
\usepackage{float}

\begin{document}


\title{Phase-dependent dissipation and supercurrent of a graphene-superconductor ring under microwave irradiation}


\author{Ziwei Dou\textsuperscript{1}, Taro Wakamura\textsuperscript{1} , Pauli Virtanen\textsuperscript{2}, Nian-Jheng Wu\textsuperscript{1,3} , Richard Deblock\textsuperscript{1} , Sandrine Autier-Laurent\textsuperscript{1} , Kenji Watanabe\textsuperscript{4} , Takashi Taniguchi\textsuperscript{4}, Sophie Gu\'{e}ron\textsuperscript{1}, 
\\H\'{e}l\`{e}ne Bouchiat\textsuperscript{1} , and Meydi Ferrier\textsuperscript{1}}
\affiliation{1. Universit\'e Paris-Saclay, CNRS, Laboratoire de Physique des Solides, 91405, Orsay, France. 
\\2. Department of Physics and Nanoscience Center, University of Jyv\"{a}skyl\"{a}, Finland 
\\3. Universit\'e Paris-Saclay, CNRS, Institut des Sciences Mol\'{e}culaires d’Orsay, Orsay, France 
\\4. National Institute for Materials Science, Tsukuba, Japan}



\begin{abstract}
A junction with two superconductors coupled by a normal metal hosts Andreev bound states whose energy spectrum is phase-dependent and exhibits a minigap, resulting in a periodic supercurrent. Phase-dependent dissipation also appears at finite frequency due to relaxation of Andreev bound states. While dissipation and supercurrent versus phase have previously been measured near thermal equilibrium, their behavior in nonequilibrium is still elusive. By measuring the ac susceptibility of a graphene-superconductor junction under microwave irradiation, we find supercurrent response deviates from adiabatic ac Josephson effect as irradiation frequency is larger than relaxation rate. Notably, when irradiation frequency further increases above the minigap, the dissipation is enhanced at phase 0 where the minigap is largest and dissipation is minimum in equilibrium. We argue that this is evidence of the nonequilibrium distribution function which allows additional level transitions on the same side of the minigap. These results reveal that phase-dependent dissipation is more sensitive than supercurrent to microwave irradiation, and suggest a new method to investigate photon-assisted physics in proximitized superconducting system.
\end{abstract}


\maketitle


\indent A junction with two superconductors coupled by a normal metal hosts Andreev bound states (ABSs), which shuttle the Cooper pairs from one superconducting bank to the other and whose energy spectrum depends on their phase difference ($\varphi$). The phase-dependent Andreev levels give rise to supercurrent $I_s$ periodic in $\varphi$, and the measurement of $I_s(\varphi)$, or the current-phase relation (CPR), has previously revealed, for example, the singlet/doublet transition in carbon nanotube quantum dots \cite{kouwenhovenQDSNS,Raphaelleprb0pi} and the helical edge states in topological materials \cite{MuraniBi,ChuanTopoCPR}. If $\varphi$ acquires a time-dependent ac component ($\delta\varphi$), generated for example by an ac magnetic field in a ring geometry, finite-time relaxation of ABSs towards equilibrium causes delay in the current response and consequently a counter-intuitive dissipation appears \cite{averinIman}. Such dissipation involves two important mechanisms: One is the relaxation of thermally excited ABSs via inelastic scattering; the other is inter-level transitions induced by microwave photons \cite{BastienPRB,BastienPRL,PauliPRBac}. In the weakly driven regime where $\delta\varphi\ll\pi$, the dissipation shed further light on the properties of the Andreev levels, revealing, for example, protected level crossing in topological junctions \cite{FuKane,MuraniMicrowave}. Strong driving power can significantly modify the distribution function from thermal equilibrium, activating additional level transitions \cite{Zaikinbook, KLAPWIJKreview}. However, this nonequilibrium state has so far mainly been investigated in dissipationless supercurrent response, finding enhanced critical current \cite{Zaikinbook,KLAPWIJKreview} or modified CPR \cite{marco, StrunkMicrowave}. Here we present the evolution of both the CPR and dissipation under microwave irradiation with different frequency and power, extracted simultaneously from the ac magnetic susceptibility of a phase-biased graphene-superconductor ring in long diffusive regime. The Andreev spectrum is characterized by a minigap whose size relative to the irradiation frequency plays an important role in the junction's response \cite{likharevSNSRev, Zhou1998}. Graphene is chosen since it has lower density of states while maintaining similar mean-free path compared to conventional normal metals \cite{chuanGraphene}. This gives access to similar minigap with a reduced supercurrent and the screening effect, enabling accurate measurement over a complete phase period \cite{BastienPRB}. 

\begin{figure}[h]
  \centering
    \includegraphics[width= 0.5\textwidth]{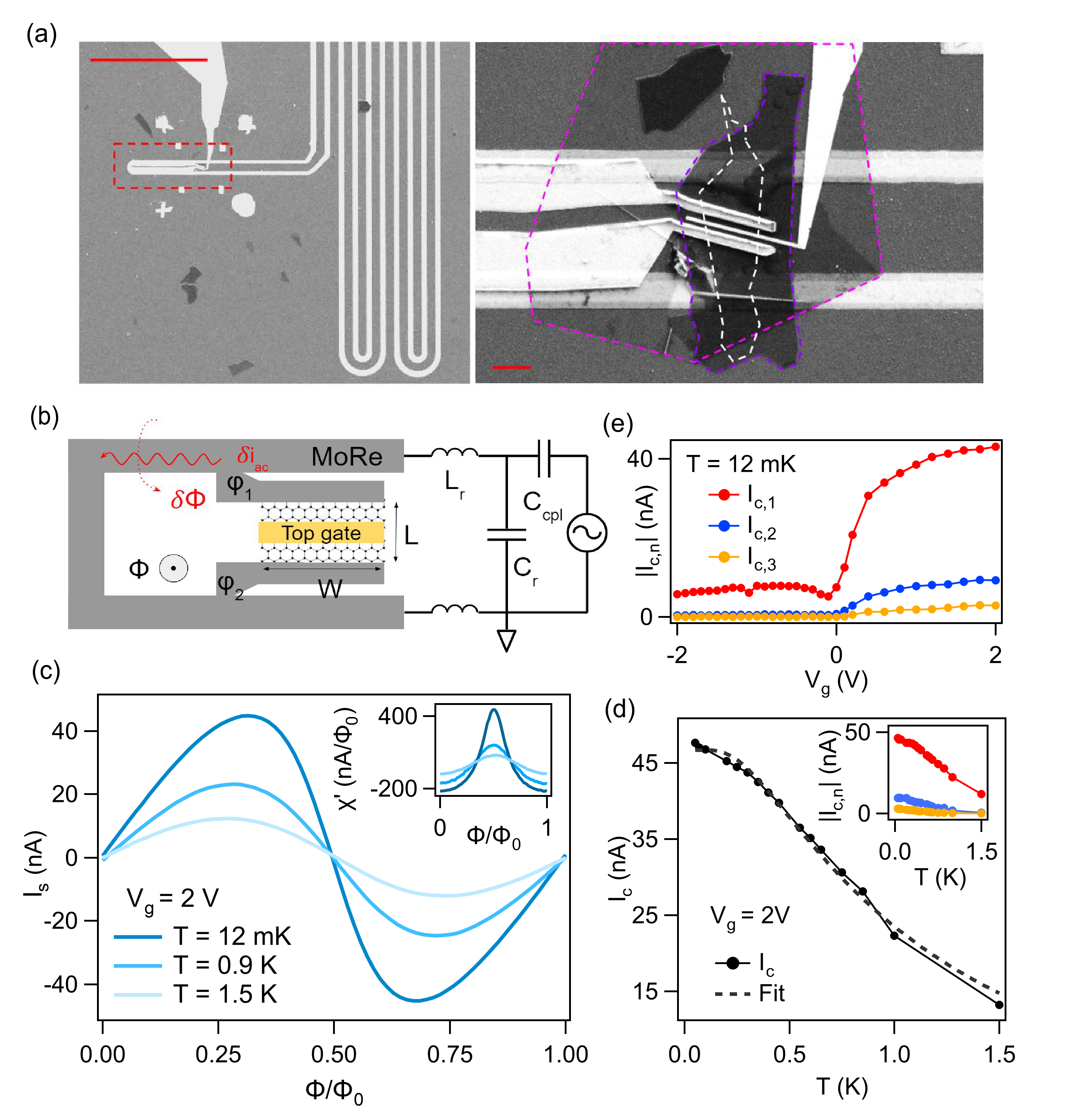}
\caption{CPRs without microwave irradiation: (a) SEM images of the device. Left: A section of the MoRe resonator. Scale bar: 100 \textmu m. Right: zoomed-in image of the SGS junction. Scale bar: 3 \textmu m. Purple and magenta dashed outlines: top and bottom BN. White dashed outline: graphene. (b) Circuit diagram. (c) CPRs at $V_g$ = 2 V, and T = 12 mK, 0.9 K and 1.5 K. Inset: $\chi'(\Phi)$. (d) The critical current $I_c$ versus $T$ at $V_g$ = 2 V. Dashed line: fitting to diffusive model with $E_{Th}$ = 34 \textmu eV. Inset: The amplitudes of the first three Fourier coefficients of the CPRs $|I_{c,n}|$ (n = 1, 2, 3) versus $T$. (e) $|I_{c,n}|$ (n = 1, 2, 3) versus $V_g$ at $T$ = 12 mK.}
\end{figure}

\indent The ac susceptibility is measured by coupling a graphene/superconductor ring to a superconducting resonator. Fig. 1(a) shows the scanning electron microscopy (SEM) images of the device. The left image shows a section of the resonator (the meander lines) made by e-beam lithography and sputtered molybdenum-rhenium (MoRe) on the undoped silicon substrate. The boron-nitride/graphene/boron-nitride (BN/G/BN) stack is fabricated using exfoliated flakes and is then connected to MoRe via side contacts \cite{graphene1d}. The details of the fabrication is given in \cite{supp}. The right image is a zoom of the junction region. The junction width is $W$ = 5 \textmu m and the length $L$ = 950 nm. The Ti/Au top-gate covers 330 nm of the total graphene length.  In Fig.1(b), the resonator is designed to have $C_{r}$ = 180 pF and $L_{r}$ = 40 nH. The loop at the end of the superconducting lines provides a coupling inductance $L_{c}$ = 355 pH. To maintain sufficiently high quality factor $Q$, the resonator is coupled to the rf coaxial cables on the dilution refrigerator through a coupling capacitor $C_{cpl}$  = 5.6 pF. At $T$ = 12 mK, the resonance frequency $\nu_r$ = 60 MHz and $Q\sim$ 300. The dc flux $\Phi$ (or the dc phase $\varphi$ = $\varphi_1$ - $\varphi_2$ = 2$\pi\Phi$/$\Phi_0$) is set by the dc magnetic field through the area defined by the ring, and is added to the ac flux $\delta\Phi$ inducing the ac current $\delta i_{ac}$. The ac susceptibility is thus defined as $\chi = \delta i_{ac}/\delta\Phi$ \cite{BastienPRB}. The rf power is heavily attenuated so that $\delta\Phi\ll\Phi_0$ and $\chi$ does not depend on the rf power. On resonance, the real (dissipationless) and imaginary (dissipative) part $\chi'$ and $\chi''$ are linked to $\nu_r$ and $Q$ via $\chi' = -2(L_{r}/L_{c}^{2})(\delta \nu_r/\nu_r)$ and $\chi'' = (L_{r}/L_{c}^{2})\delta\left(1/Q\right)$\cite{BastienPRB}. As $\Phi$ is swept, $\delta \nu_r$ and $\delta (1/Q)$ are simultaneously measured using a phase-locked feedback loop which maintains the resonator on resonance \cite{BastienPRB,franchesca,Reulet1995}. At sufficiently low frequency 60 MHz where $\chi''/\chi'\ll 1$ (justified in \cite{supp}),  $\chi'(\Phi)\approx \partial I_{s}/\partial\Phi = (2\pi/\Phi_{0})(\partial I_{s}/\partial\varphi)$ where $I_s$ is the supercurrent \cite{BastienPRB}. Thus, integrating the measured $\chi'(\Phi)$ yields $I_{s}(\varphi)$ of the junction. 

We first explore the junction without microwave irradiation. Fig. 1(c) displays the CPRs $ I_{s}(\Phi)$ at $T$ = 12 mK, 0.9 K and 1.5 K [$\chi^{'}(\Phi)$ in inset]. From the CPRs we extract the critical current $I_{c}$ as well as its first three Fourier coefficients $I_{c,n}$ (n = 1,2,3), where $I_{s}(\varphi) \approx I_{c,1}\sin(\varphi)+I_{c,2}\sin(2\varphi)+I_{c,3}\sin(3\varphi)$. $I_{c}(T)$ and $|I_{c,n}|$ are plotted in Fig.1(d). Note $I_{c,2}$ is negative and $|I_{c,2}| = -I_{c,2}$. At T = 12 mK, $|I_{c,2}|$ and $|I_{c,3}|$ have non-negligible values, consistent with the skewed CPR in Fig. 1(c) while at $T$ = 1.5 K only $|I_{c,1}|$ is dominant, meaning the CPR is sinusoidal. Assuming long diffusive junction model at  low temperature, we can fit $I_{c}(T)$ by $ I_c =  (7.7E_{Th}/eR)\left[1-1.3\exp\left(-7.7E_{Th}/3.2k_B T\right)\right]$, where Thouless energy $E_{Th}$ and normal resistance $R$ are two fitting parameters \cite{dubos}. The fitting yields $R$ = 5.6 k\textOmega { }and $E_{Th}$ = 34 \textmu eV (equivalently 400 mK or 8 GHz). The superconducting gap of MoRe is $\Delta$ = 1.8$k_B T_c$ where $T_c \approx 6 K$, thus $\Delta/E_{Th} \approx$ 20. Since $E_{Th} = \hbar v_F l_{e}/2L^2$ \cite{dubos}, the mean-free path is $l_{e} \approx$ 100 nm. The minigap $E_g = 2\times 3.1E_{Th} =$ 206 \textmu eV (or 46 GHz) \cite{minigap3Eth} is higher than the irradiation frequency accessible in later experiment. However, the above estimation assumes perfect interface and strongly diffusive transport. By numeric simulation \cite{supp}, we show that $E_g = 2\times 3.1E_{Th}$ overestimates the minigap in junction with imperfect contacts and weak disorder, which is plausible in BN-encapsulated graphene sample. The actual $E_g$ may thus fall within the energy range probed in the experiment. In \cite{supp}, we also discuss the possibility of describing the data without irradiation with a ballistic model and find similar levels of agreement . However, the data with irradiation agree better with the diffusive model. 

Fig.1(e) shows $|I_{c,n}|$ versus $V_g$ at $T$ = 12 mK. At $V_g$ = 2 V, the supercurrent is almost saturated at 44 nA, smaller than in previous studies \cite{goswami, FinkelsteinGraphene, FinkelsteinBallis} possibly because of  insufficiently filtered radiation from environment and the higher normal resistance of graphene in the un-gated regions. Meanwhile, $|I_{c,1}| \sim 5|I_{c,2}|$  at $V_g$ = 2 V, similar to a uniformly gated sample \cite{goswami}, meaning the transparency between the MoRe and graphene is still enough to preserve $I_{c,2}$. For negative $V_g$, $|I_{c,1}|$ is much reduced and $|I_{c,2}|$, $|I_{c,3}|$  become negligible, consistent with the formation of pn junctions \cite{goswami, FinkelsteinBallis}. Throughout the paper we focus on the high electron-doping regime with $V_{g}$ = 2 V for the highest susceptibility signal. 


\begin{figure}[h]
  \centering
    \includegraphics[width= 0.5\textwidth]{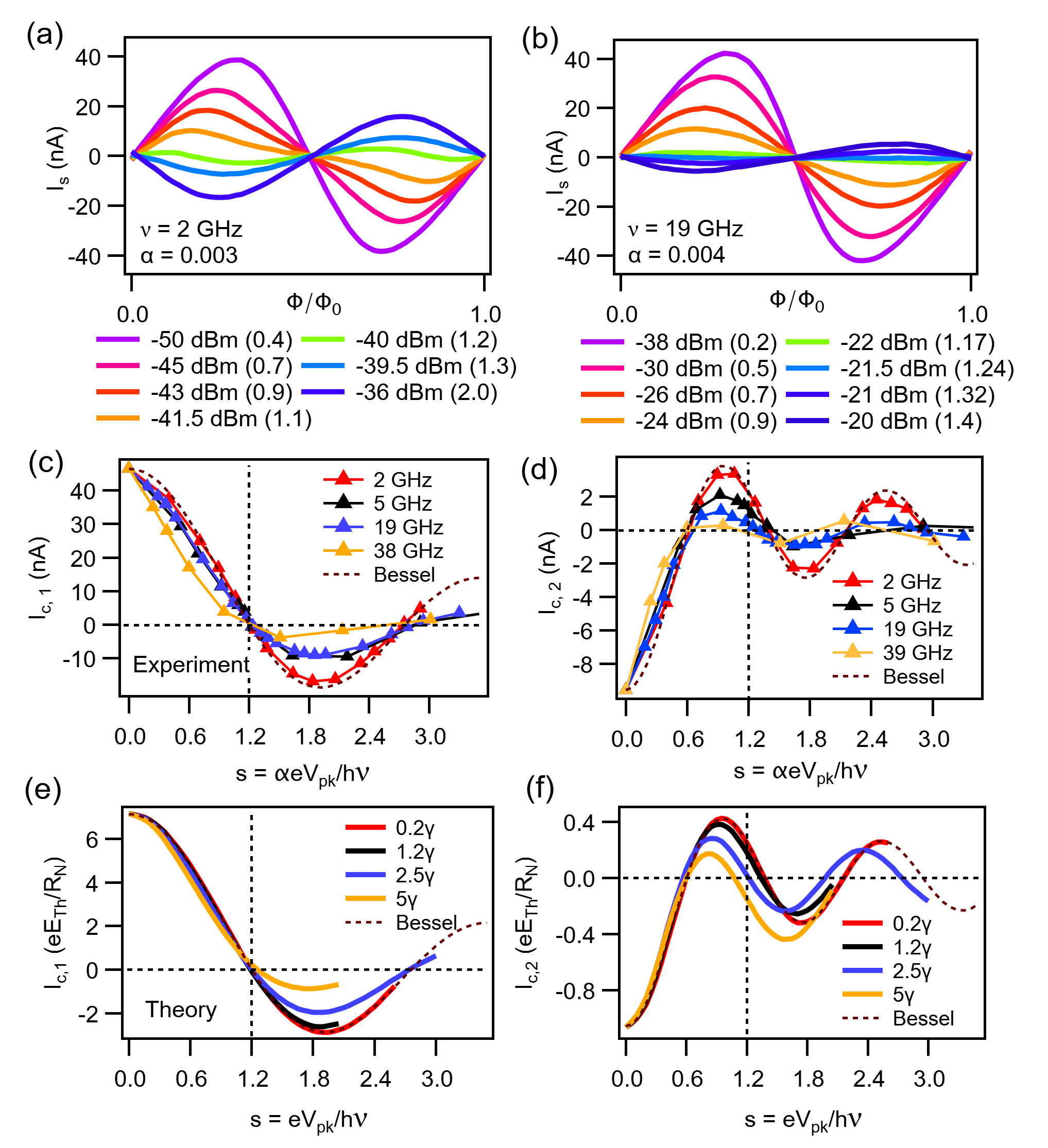}
\caption{Effects of microwave irradiation on CPR Fourier coefficients: (a) CPRs with irradiation $\nu$ = 2 GHz. The irradiation power is converted to the normalized value $s = \alpha eV_{pk}/h\nu$ in parenthesis (see text). The sign reversal of $I_s$ is observed at $s \sim$ 1.2 with halved periodicity. (b) CPRs with $\nu$ = 19 GHz. The sign reversal of $I_s$ is also seen at $s \sim$ 1.2, but without halved periodicity. (c, d) $I_{c,1}(s)$ and $I_{c,2}(s)$. All data are taken at $V_g$ = 2 V and $T$ = 12 mK. (e, f) Calculated $I_{c,1}(s)$ and $I_{c,2}(s)$ from Usadel equations with finite relaxation rate $\gamma = 1.2E_{Th}$.  $k_BT = 0.004\gamma$. $\Delta/E_{Th} = 50$. Bessel functions $I_{c,1}(0)J(s)$ and $I_{c,2}(0)J(2s)$ are plotted for (c, e) and (d, f) respectively (dashed lines). $J$ is the zero-th order Bessel function of the first kind. $I_{c,1}(0)$ = 44 nA, and $I_{c,2}(0)$ = -9 nA are the non-irradiated values from Fig. 1.}
\end{figure}

\indent After exploring the junction without irradiation and finding the CPRs agree with previous experiments \cite{goswami,FinkelsteinBallis,Bretheau2017}, we now present the data with irradiation. In Fig. 2(a, b), the CPRs are measured at two irradiation frequencies $\nu$ = 2 GHz and $\nu$ = 19 GHz, respectively. The power noted in the figure is converted to the normalized power $s = \alpha eV_{pk}/h\nu$  ($V_{pk}$ is the peak voltage at the source and $\alpha$ includes the attenuation factor from source to junction). In both figures, a sign reversal of the supercurrent occurs as the irradiation power increases. However, for $\nu$ = 2 GHz, a strong second harmonic (halved periodicity) is observed during the reversal while almost no phase-dependence is seen for $\nu$ = 19 GHz in a similar situation. In Figs. 2(c, d) we plot the power dependence of the first two Fourier coefficients of the CPRs for $\nu$ between 2 GHz and 39 GHz. At low irradiation frequency, the junction shows the adiabatic ac Josephson effect \cite{tinkham2004introduction} where ABSs follow instantaneously the oscillating $\delta\varphi$ and $I_{c,n}$ shows the Bessel function dependence. The experimental $\alpha$ is hard to calibrate accurately, and for each $I_{c,1}(s)$, $\alpha$ is chosen such that its first zero coincides with $s$ = 1.2 (the first zero of the Bessel function). The same $\alpha$ is then used for $I_{c,2}(s)$ of the respective frequency. The 2 GHz data agrees well with the Bessel function, indicating the microwave drive is adiabatic \cite{PauliMicrow}. It also shows that the electronic temperature is not significantly heated by irradiation. For higher frequencies, the agreement is less satisfactory. In particular, $I_{c,2}$ at $s = 1.2$ decreases for higher $\nu$, which is consistent with the disappearance of halved periodicity in the CPR for $\nu$ = 19 GHz. Using time-dependent Usadel equations incorporating a finite inelastic scattering rate $\gamma$ \cite{PauliMicrow}, we calculate $I_{c,1}(s)$ and $I_{c,2}(s)$ at low temperature and $\nu$ lower and higher than $\gamma$. The results are plotted in Figs. 2(e, f). At $\nu = 0.2\gamma$, $I_{c,1}(s)$ and $I_{c,2}(s)$ follow the Bessel function as expected, while for $\nu > \gamma$ they show deviation in qualitative agreement to Figs. 2(c,d). Comparing the theory ($1.2\gamma$ and $2.5\gamma$ curves) with the experiment (5 GHz and 19 GHz curves), the experimental $\gamma$ can be estimated as between 5/1.2 GHz = 4 GHz and 19/2.5 GHz = 8 GHz. Thus, $h\gamma \lesssim E_{Th}$, reasonable for SNS junctions \cite{BastienPRB}.

\begin{figure}[h]
  \centering
    \includegraphics[width= 0.5\textwidth]{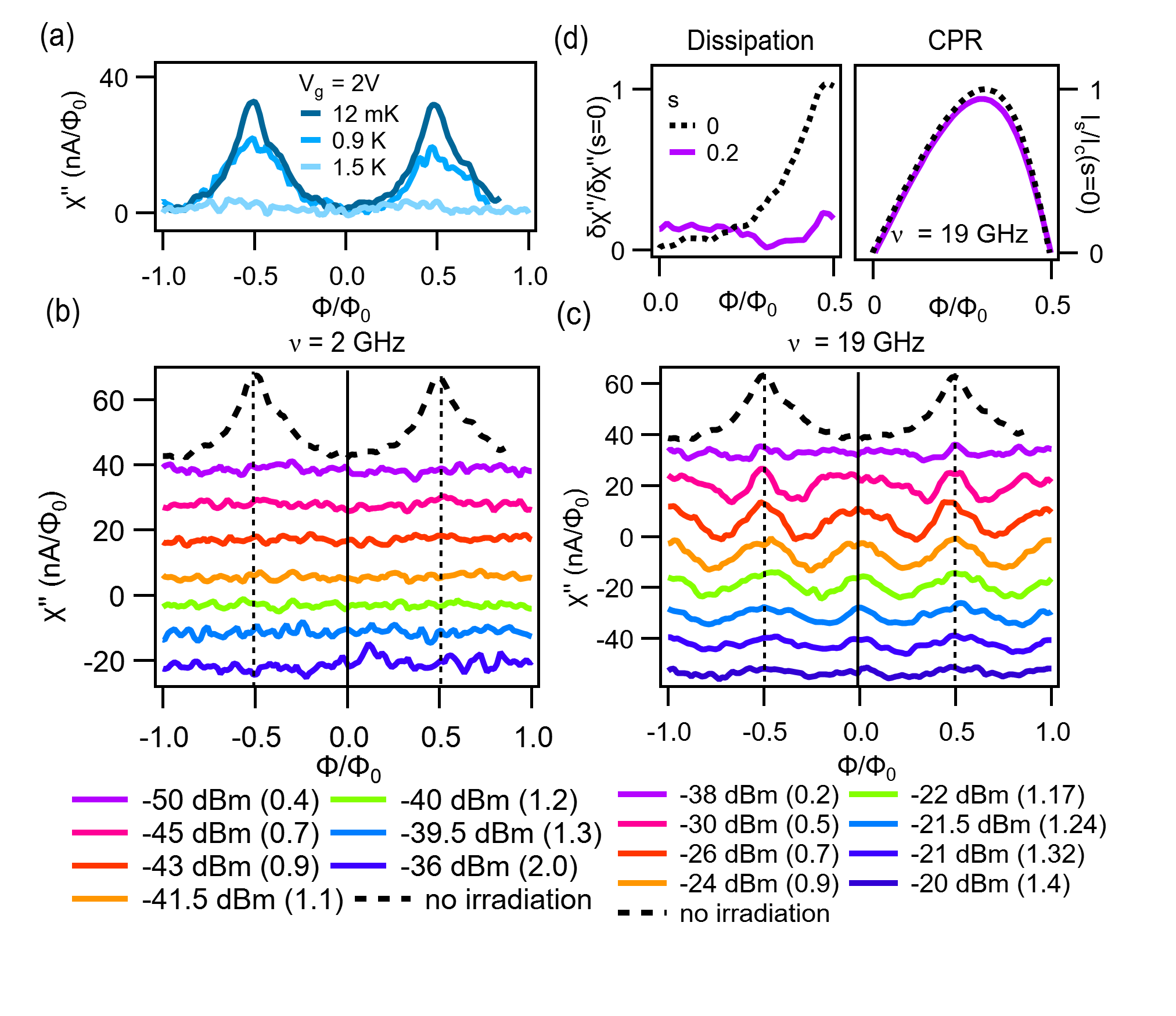}
\caption{Effects of microwave irradiation on dissipative part of the susceptibility: (a) $\chi''(\Phi)$ without irradiation at $T$ = 12 mK, 0.9 K, and 1.5 K, taken simultaneously as $\chi'(\Phi)$ leading to the CPRs in Fig. 1(c). (b) $\chi''(\Phi)$ taken simultaneously as the CPRs in Fig. 2(a). The irradiation frequency $\nu$ = 2 GHz. The irradiation power is converted to the normalized value $s$ as in Fig. 2. The black dashed line is the $\chi''(\Phi)$ without irradiation. $\chi''(\Phi)$ shows no significant $\Phi$ dependence under irradiation. (c) $\chi''(\Phi)$ taken simultaneously as the CPRs in Fig. 2(b). $\nu$ = 19 GHz. $\chi''(\Phi)$ changes from 2$\pi$-periodic with no irradiation to $\pi$-periodic under higher irradiation power. (d) Left: $\delta\chi''$ normalized by un-irradiated value ($s$ = 0). Right: $I_s$ in Fig. 2(b) normalized by critical current at $s$ = 0. $V_g$ = 2 V and $T$ = 12 mK for (b, c, d). The $\chi''$ curves are shifted vertically for clarity.}
\end{figure}

\indent The finite $\nu_r/\gamma$ gives rise to a nonzero dissipation $\chi''$ \cite{BastienPRB}. Fig. 3(a) shows $\chi''(\Phi)$ taken simultaneously as the CPRs in Fig. 1(c) without irradiation. $\chi''(\Phi)$  peaks at $0.5\Phi_0$ and its height decreases with temperature. This higher dissipation is a result of the minigap closing which allows more excitation-relaxation events between Andreev states \cite{BastienPRB}. At low temperature, $\delta\chi''/\delta\chi' \sim 20$ [here $\delta\chi = \chi(0.5\Phi_0)-\chi(0)$] is indeed similar to $\nu_r/\gamma$ estimated independently from CPRs, as discussed in \cite{supp}. Figs. 3(b, c) show $\chi''(\Phi)$ under irradiation taken simultaneously as the CPRs in Figs. 2(a, b). For both $\nu$ = 2 GHz and 19 GHz, the small irradiation power flattens $\chi''(\Phi)$. More data in \cite{supp} shows a gradual diminution in $\delta\chi''$  for smaller power. This is compared with the CPR, as illustrated in Fig. 3(d): $\delta\chi''(s = 0.2)$ is decreased by 80\% of the un-irradiated $\delta\chi''(s = 0)$ whereas the CPR is almost unaffected, demonstrating the much higher power sensitivity of $\chi''$ than CPR. At high irradiation power, the dissipation response is drastically different for $\nu$ = 19 GHz: Instead of the flat $\chi''(\Phi)$ in Fig. 3(c), an additional lobe appears in Fig. 3(d) around flux $0$ and the peak at $0.5\Phi_0$ becomes pronounced again, which turns $\chi''(\Phi)$ into a function quasi-periodic in $\Phi_0/2$.  To our knowledge, this emergence of an enhanced dissipation at flux 0 under high irradiation frequency and power has never been reported and requires an explanation.

 \begin{figure}[h!]
  \centering
    \includegraphics[width= 0.5\textwidth]{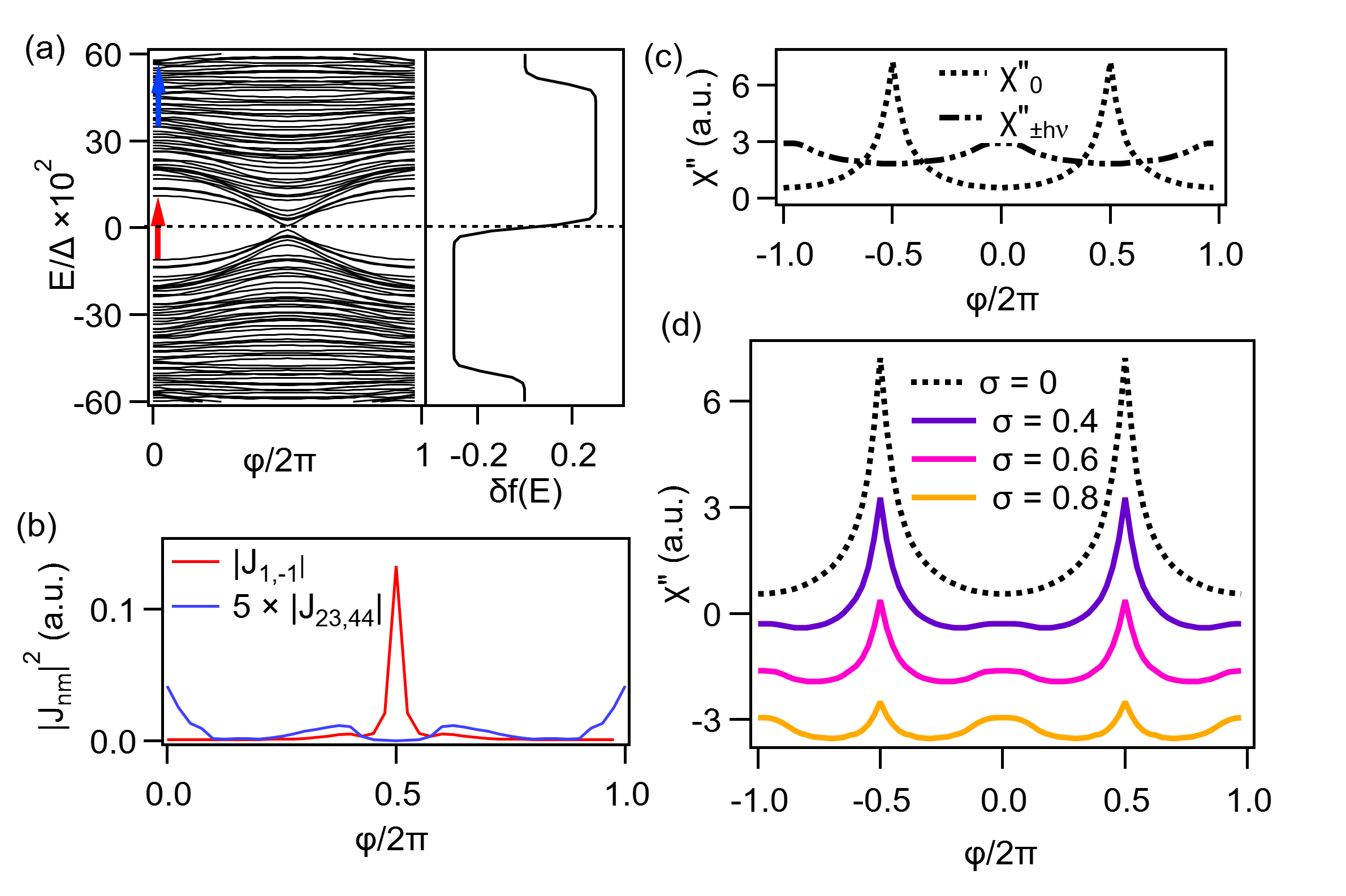}
\caption{Calculation of $\chi''(\Phi)$ with nonequilibrium distribution function: (a) Andreev spectrum of a long diffusive system displaying a minigap $E_g \sim 0.2\Delta$. $\Delta$ is the superconducting pairing potential. Side plot: distribution function change $\delta f(E)$ due to irradiation (see text). (b) The interband term $|J_{1,-1}|^2$ (red) and the intraband term $|J_{23,44}|^2$ (blue). The corresponding transitions are marked by the arrows in the same colors in (a). (c) $\chi''_{0}(\varphi)$ calculated with $f_{FD}(E)$ (dashed line) and $\chi''_{\pm h\nu}(\varphi)$ calculated with $f_{FD}(E\pm h\nu)$ (dashed-dotted line). (d) Total $\chi''(\Phi)$. $\sigma$ stands for the irradiation power ($\sigma$ = 0 being no irradiation).  Curves are offset for clarity. Similar to experiment, $k_BT= h\nu_r = 0.01\Delta\ll E_g$, relaxation rate $h\gamma = 0.15\Delta \sim 0.5E_g$ and irradiation frequency $h\nu = 0.5\Delta \sim 2E_g$.  System size: $W\times L = 50\times30$ sites.}
\end{figure}

While $\chi''(\Phi)$ with nonzero $\nu_r$ can be calculated by similar Usadel equation approach \cite{PauliPRBac} to that in Fig. 2 for the CPRs in the dc limit, it is not straightforward to include both the low-frequency ac flux and the high-frequency irradiation. In order to explain Fig. 3(c), we thus turn to the Kubo formalism, which has well described the $\chi''(\Phi)$ of diffusive junctions in previous experiments without irradiation \cite{Trivedi,BastienPRB,meydi}. In our junction, the rf frequency used for susceptibility measurement $\nu_r$ is much smaller than the inelastic scattering rate $\gamma$, thus $\chi''$ can be approximated by \cite{supp}:

\begin{equation}
\label{eqn:Kubofull}
\chi'' \approx -\operatorname{Im}\left[\displaystyle\sum_{n, m \neq n} |J_{nm}|^2\frac{f_n-f_m}{\epsilon_n-\epsilon_m}\frac{ih\nu_r}{i(\epsilon_n-\epsilon_m-h\nu_r)+h\gamma}\right]
\end{equation}

\noindent where $J_{nm} = (e\hbar/m_ei)\langle n|\nabla|m\rangle$ is the off-diagonal element of the current operator, $\epsilon_n$ is the n-th Andreev level. The effect of the microwave irradiation is included phenomenologically in the distribution function:

\begin{equation}
\label{eqn:FD}
f_n=(1-\sigma)f_{FD}(\epsilon_n)+\sigma/2[f_{FD}(\epsilon_n+h\nu)+f_{FD}(\epsilon_n-h\nu)]
\end{equation}

\noindent where $\nu$ is the irradiation frequency and $\sigma$ stands for the irradiation power. Without irradiation ($\sigma$ = 0), $f_n$ follows the Fermi-Dirac distribution $f_{FD}(\epsilon_n) = 1/[1+\exp(\epsilon_n/k_BT)]$. With the irradiation, the change in distribution function from the equilibrium one $\delta f(\epsilon_n) = f_n - f_{FD}$ shown in Fig. 4(a) qualitatively reproduces the results based on the time-dependent Green's function approach near $\varphi$ = 0 \cite{PauliMicrow,Gunsenheimer_1998,EschrigMicrow}: The Andreev state occupation is depleted (enhanced) within $h\nu$ below (above) the minigap by irradiation photons. The total $\chi''$ can thus be rewritten as $\chi'' = (1-\sigma)\chi''_0+\sigma/2[\chi''_{+h\nu}+\chi''_{-h\nu}]$, where $\chi''_0$ and $\chi''_{\pm h\nu}$ are calculated with Eq.\ref{eqn:Kubofull} using $f_{FD}(E)$ and $f_{FD}(E\pm h\nu)$ respectively. We numerically calculate these terms in a diffusive SNS junction using the tight-binding method \cite{kwant, meydi}. See details in \cite{supp}. The Andreev spectrum near $E$ = 0 is plotted in Fig. 4(a) showing a minigap. Fig. 4(b) displays two examples $|J_{1,-1}|^2$ and $|J_{23,44}|^2$ whose corresponding $(f_n-f_m)/(\epsilon_n-\epsilon_m)$ are nonzero. They represent ``interband" transition [red arrow] and ``intraband" transition [blue arrow], respectively. As $\sigma$ goes from 0 to 0.8, $\chi''(\varphi)$ evolves from a 2$\pi$-periodic function to a quasi $\pi$-periodic function, which is the key feature in Fig. 3(c). By comparing Figs. 4(c) and (d), the rising lobe at phase 0 is attributed to the nonequilibrium terms $\chi''_{\pm h\nu}$ and can be understood as follows: At low temperature, the equilibrium $f_{FD}$ is close to a step function, thus only interband terms $|J_{n,-n}|^2$ contribute to Eq.(6). Since $|J_{n,-n}|^2$ generally peaks at $\pi$ (e.g. $|J_{1,-1}|^2$) \cite{meydi}, $\chi''_{0}(\pi)$ is also high. Meanwhile, under irradiation with $h\nu > E_g$, the distribution function is modified and states near the minigap are partially occupied, enabling more intraband transitions. The intraband terms $|J_{nm}|^2$, which in general have higher magnitudes around phase 0 (see $|J_{23,44}|^2$), produces an enhanced $\chi''_{\pm h\nu}(0)$. The same simulation with $h\nu < E_g$ only shows the $2\pi$-periodic $\chi''$. The flat $\chi''(\Phi)$ at small power [$s$ = 0.2 in Fig. 3(c)] is not captured by our simple model. However, this may be due to the detailed form of the distribution function deviating from Eq.\ref{eqn:FD}. 

\indent In conclusion, we measured the ac susceptibility of a phase-biased graphene/superconductor junction in response to the irradiation of microwave photons. For low irradiation frequency, the power dependence of the CPR Fourier coefficients follows the adiabatic response, while the phase dependance of the dissipation almost disappears at small irradiation power, demonstrating its higher power sensitivity than supercurrent. At higher irradiation frequency, the CPR responses deviate from adiabaticity and agree with the quasiclassical theory including finite relaxation rate. More remarkable effects manifest in dissipation as the irradiation frequency further increases above the minigap: The dissipation is enhanced around phase 0 which results in a $\pi$-periodic phase dependence. We argue that this new phenomenon happens as the system is driven strongly out of equilibrium. The phase-dependent intraband transitions, which are highly suppressed in equilibrium at low temperature, thus become possible with a nonequilibrium distribution function induced by microwave irradiation. The measurement exemplified here may lead to a more detailed understanding of other nonequilibrium physics in proximitized superconducting system \cite{watfa, deacontopo, noneqFluc}. It also offers insight to the development of novel superconducting quantum devices operating in microwave field \cite{steeleGmicroloss, Leebolometer2020, Kokkoniemi2020}. 

\begin{acknowledgments}
We acknowledge insightful discussion with T. T. Heikkil\"{a} (University of Jyv\"{a}skyl\"{a}, Finland), M. Aprili, and J. Est\`{e}ve (Universit\'e Paris-Saclay, France). We also acknowledge financial support from ERC 833350, Japan Society for the Promotion of Science (2017-684), Labex PALM, and ANR JETS (ANR-16-CE30-0029-01).
\end{acknowledgments}

%

\end{document}



\title{Supplementary Materials: Supercurrent and dynamic dissipation of a phase-biased graphene-superconducting
junction under microwave irradiation}







\maketitle

\section{Device fabrication}
The superconducting meander lines are defined by e-beam lithography using chemical semi-amplified positive e-beam resist (CSAR) with $\sim$ 200 nm thickness for high spatial resolution \cite{csar}, on a high-resistivity SiO\textsubscript{2}/Si substrate. Polycrystalline molybdenum-rhenium (MoRe) alloy is deposited by sputtering in Ar gas with thickness $\sim$ 60 nm. The graphene and BN flakes are exfoliated and identified via optical contrast on a Si substrate with 290 nm thick SiO\textsubscript{2}. The flakes are picked up by PDMS polymer to form the BN/G/BN stack which is then placed at the end of the MoRe lines \cite{Castellanos_Gomez_2014}. Another e-beam lithography defines the top-gate region which goes across the MoRe lines. The shorting between the top-gate and the MoRe line is prevented by the top BN whose thickness is $\sim$ 20 nm measured by tapping mode atomic force microscopy. The top-gate is formed by thermally evaporated titanium/gold (5 nm/50 nm). A final e-beam step defines the contacts between the MoRe lines and graphene. After reactive-ion etching through the top BN to expose the graphene edge, the sample is annealed in high vacuum at 80 \textsuperscript{o}C for 1 hour before MoRe sputtering to form the 1d contact to graphene \cite{graphene1d}. 

\section{Tight binding model for the SNS ring: Ballistic vs diffusive models}

In Bogoliubov-de Gennes formalism, the Hamiltonian of the SNS junction is written in Nambu space as:
\begin{equation}
\label{eqn:Hamil}
\mathcal{H} =  
\begin{pmatrix}
H-E_F & \Delta \\
\Delta^* & E_F-H^* \\
\end{pmatrix}
\end{equation}
\noindent where $H$ is the system without superconductivity and $E_F$ is the Fermi level. The complex pairing potential $\Delta = \Delta_0e^{i\varphi}$ incorporates the phase $\varphi$ of the superconducting electrodes. $\Delta_0$ is nonzero only in the superconducting regions. Since the data in the experiment is taken with high n-type doping, we approximate graphene lattice by 2d square lattice in the model. Discretizing Eq.\ref{eqn:Hamil} \cite{kwant} thus leads to:

\begin{equation}
\label{eqn:HamilDiscret}
\mathcal{H} = \displaystyle\sum_{n=1}^{N}[\epsilon_i\sigma_z+\Delta_0 e^{i\varphi}\sigma_x]|i\rangle\langle i|+\displaystyle\sum_{i\neq j}t_{ij}\sigma_z|i\rangle\langle j|
\end{equation}	

\noindent where $i$ represents the i-th site and N is the total site number, $\sigma_{x,z}$ are the Pauli matrices operating on Nambu space. $\epsilon_i$ and $t_{ij}$ are onsite potential of the i-th site and hopping energy between i-th and j-th sites respectively. Considering only the nearest-neighbor coupling, $t_{ij} = -t\delta_{i,j\pm1}$ ($\delta$ is kronecker function). The onsite potential $\epsilon_i = 4t - E_F + V_i$ where $V_i$ is the total electrostatic potential in the normal region. We set the Fermi level $E_F$ = 4t so that the conduction band is half-filled. In the ballistic case $V_i$ is set as 0. In the diffusive case $V_i$ is modelled by a uniformly distributed random number within the range $[-D, D]$ ($D$ is the disorder strength). The gating effect can also be included in $V_i$ by offseting the potential in the gated region. 

The Andreev spectrum is calculated by diagonalizing the Hamiltonian for each phases between 0 and $2\pi$ and the supercurrent is calculated according to \cite{likharevSNSRev}:

\begin{equation}
\label{eqn:Is}
I_s = \displaystyle\sum_n f_n \frac{\partial\epsilon_n}{\partial\varphi}
\end{equation}

\noindent where $\epsilon_n$ is the n-th Andreev spectrum and $f_n$ is the equilibrium Fermi-Dirac distribution function for $\epsilon_n$. Throughout the simulation we set $t = 1$ and $a = 1$. We also set $k_B = 1$ and $\hbar = 1$. The total length of the N region is $L = 30$, and width $W = 4L = 120$ (total site number is 3600) except for the 1d ballistic case ($W\times L = 1\times30$). 

\subsection{Ballistic model}

Assuming ballistic transport, the experimental superconducting coherence length $\xi = \hbar v_F/\Delta$ where $v_F$ = 10\textsuperscript{6} m/s in graphene and $\Delta = 1.8k_BT_c$ for MoRe ($T_c \sim$ 6 K). Thus $\xi \approx$ 700 nm $\sim 2/3L$ (The experimental $L$ = 950 nm). For tight-binding model, since $E_F = 4 - 2\cos(k_F)$ is set as 4, $v_F = \partial E/\partial k = 2\sin(k_F) = 2$. Choosing $\Delta = 0.1 \ll E_F$, thus $\xi = v_F/\Delta = 20 = 2/3L$ similar to the relative length scale in the experiment. The Andreev spectra and their corresponding current phase relations (CPRs) are computed for three cases (according to Eq. \ref{eqn:Is}) : (I) 1d ballistic case; (II) 2d ballistic case; (III) 2d ballistic case with doping inhomogeneity.

\begin{figure}[h!]
  \centering
    \includegraphics[width= 0.5\textwidth]{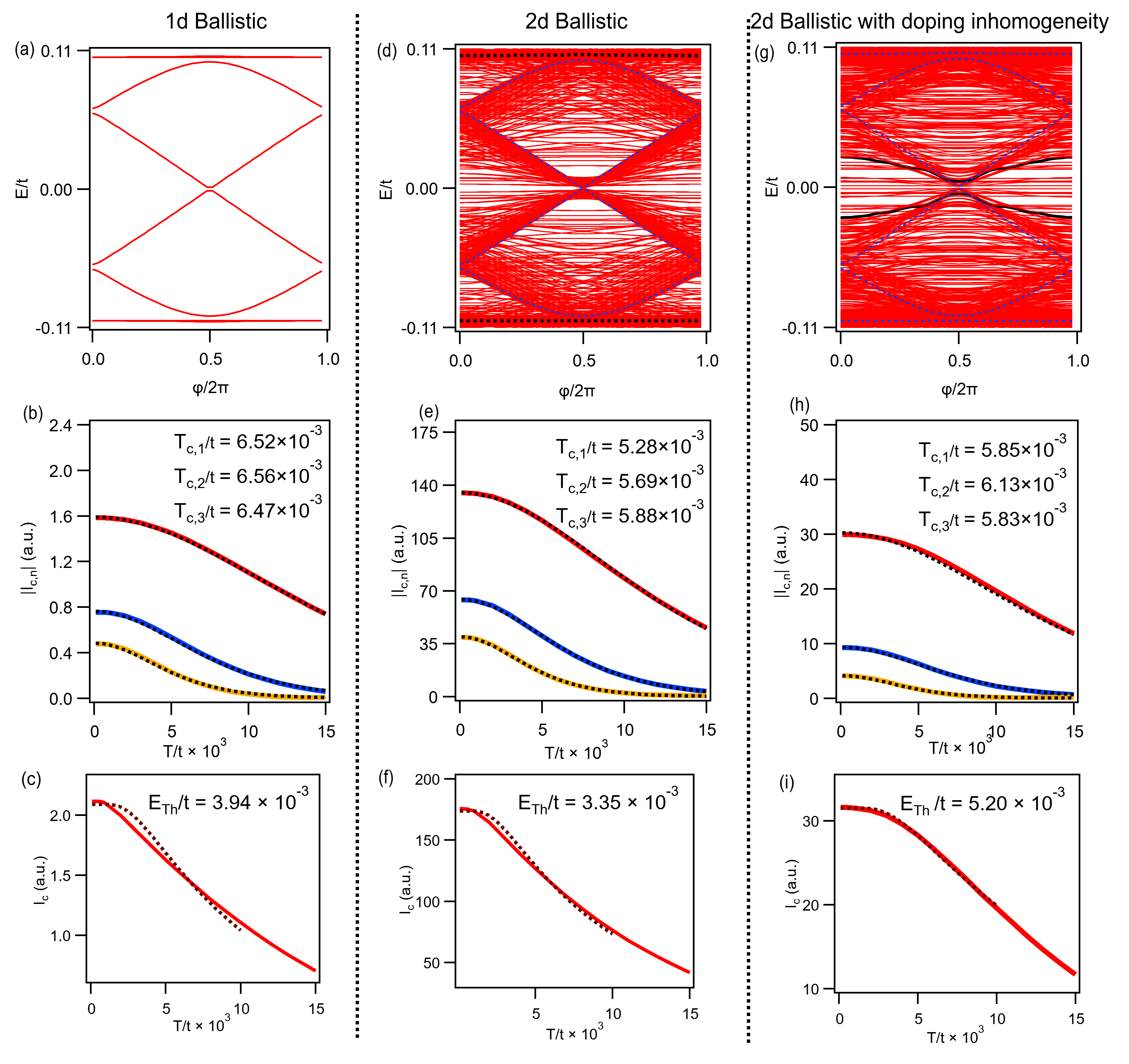}
\caption{Tight-binding simulations on ballistic system: (a) Andreev spectrum for 1d ballistic system ($W = 1$). The level spacing at $\varphi$ = 0 is $\Delta E/t$ = 0.124, close to the analytical $\Delta E/t = \pi\hbar v_F/(L+\xi)$ = 0.125. (b) $|I_{c,n}|(T) (n = 1,2,3)$ for the 1d ballistic case, fit to Eq.\ref{eqn:IcnTballis} (dashed lines). $T_{c,n}$ from the fits are listed in the graph. Analytically $T_{c,n}/t = \hbar v_F/2\pi(L+\xi) = 6.37 \times 10^{-3}$. (c) $I_{c}(T)$ for the 1d ballistic case, fit to Eq.\ref{eqn:Icdubos}. $E_{Th}$ from the fitting is listed in the graph. (d, e, f) Andreev spectrum, $|I_{c,n}|(T) (n = 1,2,3)$, and $I_{c}(T)$ for 2d ballistic system ($W = 120$). (g, h, i) Andreev spectrum, $|I_{c,n}|(T) (n = 1,2,3)$, and $I_{c}(T)$ for 2d ballistic system with doping inhomogeneity by gating  ($E_F/t = 2.5$ for the ungated normal region, close to the bottom of the conduction band $E = 2$). A soft minigap $E_g/t = 0.048$ is seen (black solid lines).}
\label{fig:balliskwant}
\end{figure}

Fig. \ref{fig:balliskwant}(a) reproduces the 1d ballistic Andreev spectrum with linear levels as previous analytical results \cite{ishii,cayssolsns, cayssolDirac}. The level spacing at $\varphi$ = 0 is $\Delta E/t$ = 0.124, close to the analytical $\Delta E/t = \pi\hbar v_F/(L+\xi)$ = 0.125 \cite{ishii,cayssolsns, cayssolDirac}. Similar to the main text, the Fourier coefficient amplitudes of CPRs $|I_{c,n}| (n = 1,2,3)$ and the critical current $I_c$ at each temperature are extracted. $|I_{c,n}|(T)  (n = 1, 2, 3)$ is plotted in Fig. \ref{fig:balliskwant}(b) and fit to the expression \cite{cayssolsns}: 

\begin{equation}
\label{eqn:IcnTballis}
|I_{c,n}|=I_{c0,n}\frac{T}{T_{c,n}}\sinh\left(\frac{T}{T_{c,n}}\right)
\end{equation}
 
\noindent where $T_{c,n} = \hbar v_F/2\pi(L+2\xi)$ and $I_{c0,n}$ is the zero temperature magnitude. The fittings almost exactly agree with the numerics. Also $T_{c,n}$ from the fittings are in good agreement with the analytical $T_{c,n} = \hbar v_F/2\pi(L+\xi)$ for all $n$ \cite{cayssolsns}. 

In order to see whether diffusive/ballistic transport regimes can be differentiated by the temperature dependence measurement of CPRs, it is illustrative to see how well the ballistic simulations can be fit to the diffusive model. In Fig. \ref{fig:balliskwant}(c), $I_{c}(T)$ is fit by the same expression used in the main text \cite{dubos}: 

\begin{equation}
\label{eqn:Icdubos}
eRI_c/E_{Th}=b\left[1-1.3exp\left(-bE_{Th}/3.2k_BT\right)\right]
\end{equation}

\noindent where the Thouless energy $E_{Th}$ and normal resistance $R$ are two fitting parameters. $b$ is a constant 7.7438 set by the ratio $\Delta/E_{Th}$. From the fit, $E_{Th} = 3.94\times 10^{-3}t$, which only accounts for 60\% of $T_{c,n}$ and is not relevant to the real energy scales in the 1d ballistic case. Also, since $E_{Th} = \hbar v_Fl_{e}/2L^2$, the mean-free path $l_{e} = 3.55 \approx 1/8L$, clearly not accurate for the ballistic system.

Extending from the 1d ballistic case to the 2d case by increasing $W$ from 1 to 120 ($6\xi$), the Andreev spectrum in Fig. \ref{fig:balliskwant}(d) shows many more levels coming from the modes whose momenta are not perpendicular to the SN interface.  Comparing with the 1d spectrum (the blue dashed lines), one finds that the 2d levels are denser around the 1d levels, similar to the quasiclassical calculation \cite{cuevasballis}. In Figs. \ref{fig:balliskwant}(e,f), by fitting $|I_{c,n}|(T) (n = 1,2,3)$ and $I_{c}(T)$ to the ballistic and diffusive expression Eq.\ref{eqn:IcnTballis} and Eq.\ref{eqn:Icdubos} respectively, $T_{c,n}$ is still close to the theoretical value $\hbar v_F/2\pi(L+\xi)$. Thus the characteristic temperature scale in 2d ballistic junction can be estimated by the 1d ballistic model. Meanwhile, $E_{Th}$ resulting from the diffusive fit is again around 60\% of $T_{c,1}$ and $l_{e} = 3.55 \approx 1/10L$, meaning the diffusive model is inaccurate in describing the 2d ballistic junction. 

Finally, by introducing doping inhomogeneity, the Andreev spectrum is plotted in Fig. \ref{fig:balliskwant}(c). Compared with the homogeneous case, a ``soft gap” (a gap with non-zero but reduced density of state) appears, whose edge is highlighted by the solid black line. Indeed, the doping inhomogeneity can act as elastic scatterer which leads to a diffusive-like Andreev spectrum. Similar features have also been reported in the local density of states measured by the tunneling spectrum in a graphene/superconductor junction \cite{Bretheau}. The fitting of $|I_{c,n}|(T) (n = 1,2,3)$ by Eq. \ref{eqn:IcnTballis} yields $T_{c,n}$ again close to 1d ballisitc theoretical value. The diffusive model, on the other hand, fits to $I_{c}(T)$ with better quality and produces $E_{Th}$ similar to $T_{c,n}$ from the ballistic model. Therefore in the inhomogeneous case, both ballistic and diffusive model yield similar characteristic energy scales and do not indicate which one agrees better with the true transport regime. Also, $l_{e} = 3.55 \approx 1/6L$ close to the length of the gate potential barrier $1/3L$.

\subsection{Diffusive model}

\begin{figure}[h!]
  \centering
    \includegraphics[width= 0.5\textwidth]{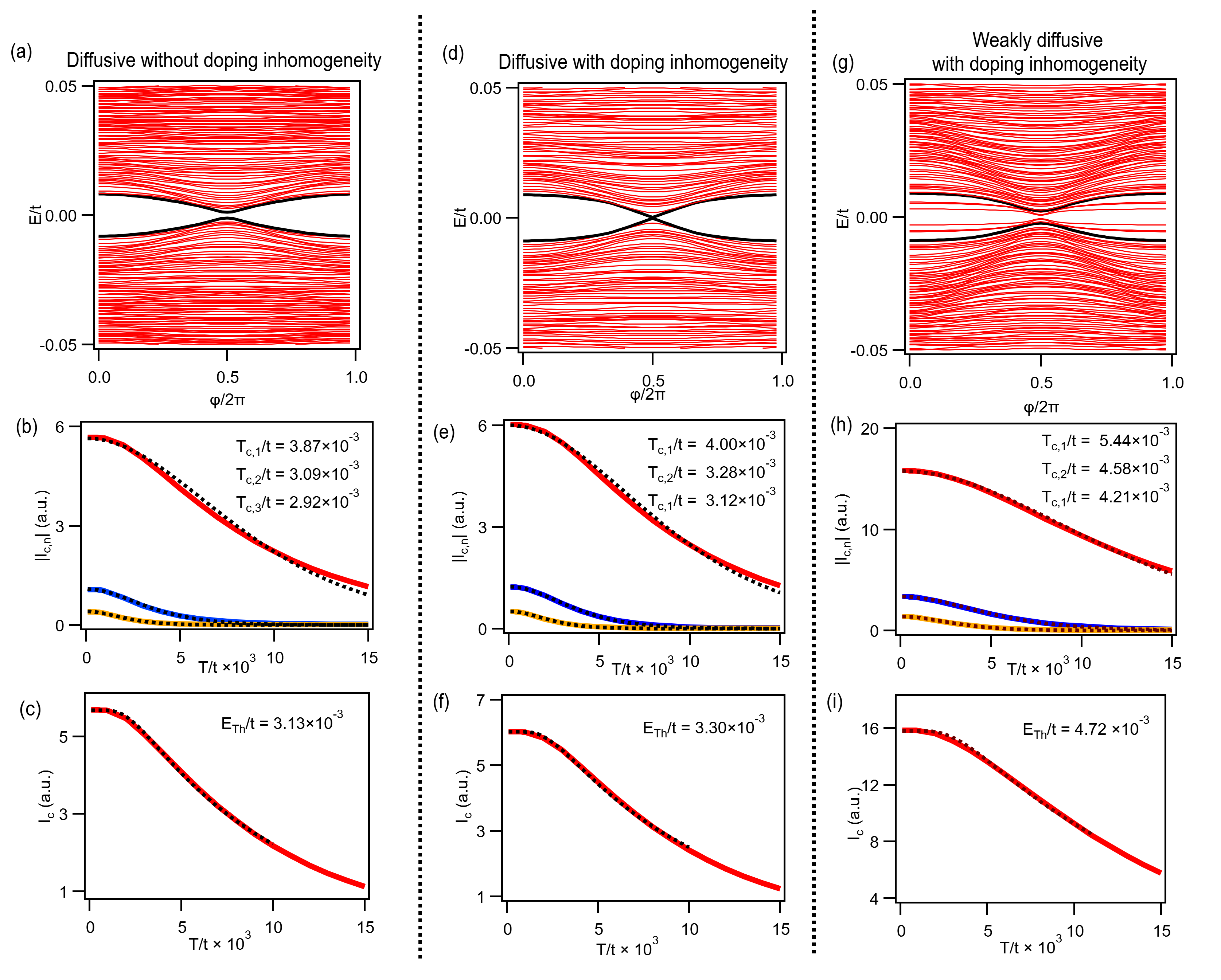}
\caption{Tight-binding simulations on diffusive system: (a) Andreev spectrum for diffusive system without doping inhomogeneity. The minigap $E_g/t = 16\times 10^{-3}$. The gap edge is marked by black solid lines. (b) $|I_{c,n}|(T) (n = 1,2,3)$ fit to Eq.\ref{eqn:IcnTballis} (dashed lines). $T_{c,n}$ from the fittings are listed in the graph. (c) $I_{c}(T)$ fit to Eq.\ref{eqn:Icdubos}. $E_{Th}$ from the fitting is listed in the graph. (d, e, f) Andreev spectrum, $|I_{c,n}|(T) (n = 1,2,3)$, and $I_{c}(T)$ for diffusive system with doping inhomogeneity. (g, h, i) Andreev spectrum, $|I_{c,n}|(T) (n = 1,2,3)$, and $I_{c}(T)$ for weakly diffusive system with doping inhomogeneity. A soft minigap $E_g/t = 18\times 10^{-3}$ is seen in the spectrum, almost twice less than $2\times 3.1E_{Th}$ estimated from $I_{c}(T)$. Lattice size $W\times L = 120\times 30$. Mean-free length: $L \sim 15 l_{e}$.}
\label{fig:diffkwant}
\end{figure}

The diffusive junction can be modeled by setting a large nonzero disorder strength $D =$ 2.5t \cite{meydi}. The Andreev spectrum is as Fig. \ref{fig:diffkwant}(a). The spectrum clearly develops a phase-dependent minigap (highlighted by the black line) which is maximum at $\varphi = 0$ and closes at $\varphi = \pi$, characteristic of the long diffusive junction. From the spectrum, the minigap $E_g = 16\times 10^{-3}$. According to the analytical model \cite{likharevSNSRev}, $E_g = 2\times 3.1E_{Th}$,  thus the Thouless energye $E_{Th} = 2.62\times 10^{-3}$. Since $E_{Th} = \hbar v_Fl_{e}/2L^2$, the mean-free path $l_{e} = 2.35$. Therefore $L \sim 15 l_{e}$ and the long diffusive junction model is indeed applicable. Similar to the ballistic case, $|I_{c,n}|(T) (n = 1,2,3)$ and $I_{c}(T)$ are plotted in Figs. \ref{fig:diffkwant}(b,c), respectively. The fittings to the ballistic and diffusive expressions Eq.\ref{eqn:IcnTballis} and Eq.\ref{eqn:Icdubos} yield similar $E_{Th}$ and $T_{c,n}$. Similar results are obtained by adding doping inhomogeneity to the diffusive system [Figs. \ref{fig:diffkwant}(d-f)]. Therefore, although we know \textit{a priori} the system is in the diffusive regime, the fittings to the supercurrent dependence on temperature alone cannot distinguish which one of the two models, ballistic or diffusive, applies better.  

In the experiment, $E_{Th}/h$ = 8 GHz and thus the minigap $E_g = 2\times 3.1E_{Th}$ = 46 GHz, much higher than the irradiation frequency (19 GHz) with which $\pi$-periodic $\chi''(\varphi)$ is observed. However, the relation $E_g = 2\times 3.1E_{Th}$ is valid only in deeply diffusive regime. Figs. \ref{fig:diffkwant}(g-i) display the Andreev spectrum, $|I_{c,n}|(T) (n = 1,2,3)$, and $I_{c}|(T)$ in a weakly diffusive junction (disorder strength $D$ = 0.5). The spectrum exhibits a soft minigap with $E_g = 18\times 10^{-3}t$ (gap edges marked by black solid lines). From $I_{c}(T)$, the diffusive model fitting yields $E_{Th} = 4.72\times 10^{-3}t$. Thus the minigap estimated by $2\times 3.1E_{Th} =  30\times 10^{-3}t$, almost twice larger than the true minigap known directly from the spectrum. Therefore, the junction in the experiment is likely to operate in the weakly diffusive regime where $E_{Th}$ obtained by $I_{c}(T)$ fitting overestimate the true minigap, thus $E_g < 2\times 3.1E_{Th}$.

\section{Comparison between ballistic and diffusive models with experiment}

\begin{figure}[h!]
  \centering
    \includegraphics[width= 0.5\textwidth]{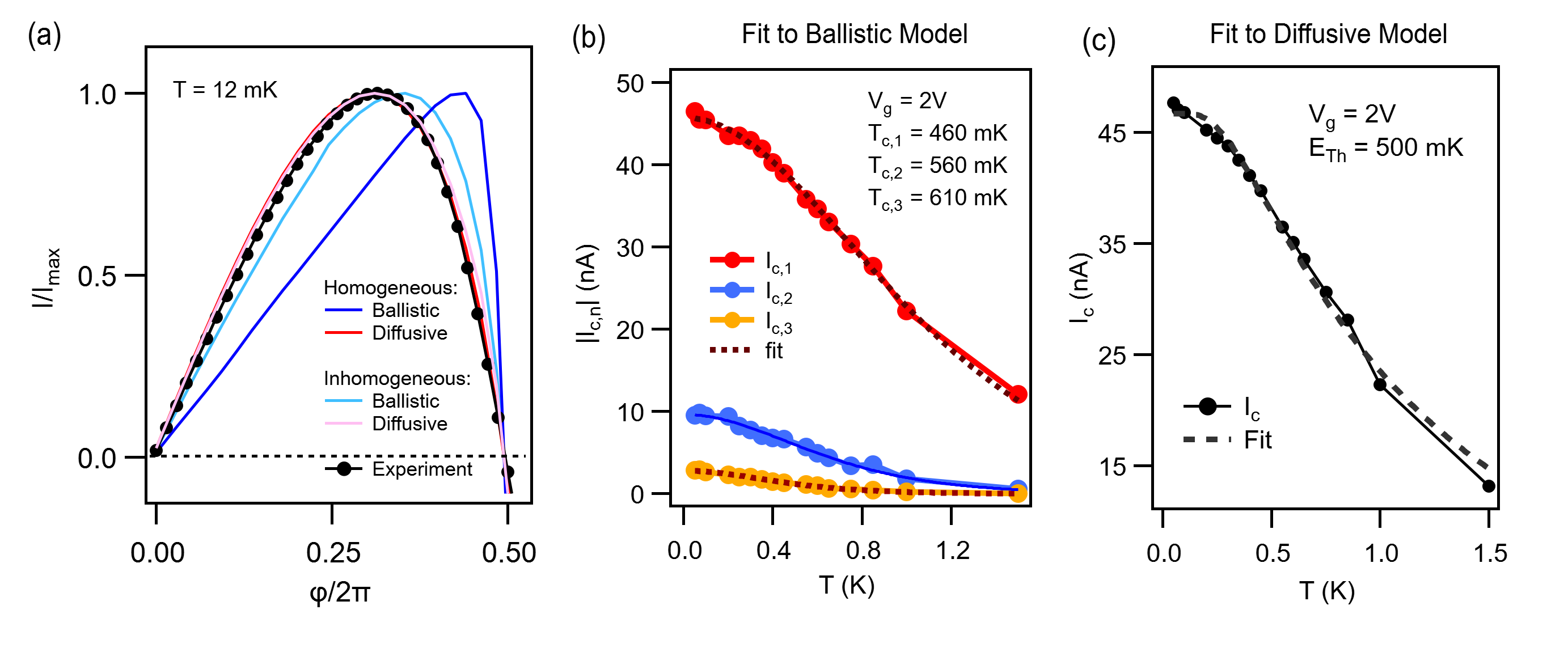}
\caption{Comparison between ballistic and diffusive model with experiment: (a) CPR normalized by the critical current $I_c$ from the experiment (black circles, $V_g$ = 2V and $T$ = 12 mK) and the calculation assuming different transport regimes marked in the graph (lines). The experimental CPR is closer to the simulated CPR using the diffusive model. (b) Experimental $|I_{c,n}|(T) (n = 1,2,3)$ fit by the ballistic model Eq.\ref{eqn:IcnTballis}. (c) Experimental $|I_{c}(T)$ fit by the diffusive model Eq.\ref{eqn:Icdubos}.}
\label{fig:expkwant}
\end{figure}

Fig. \ref{fig:expkwant}(a) compares directly the normalized experimental CPR with that calculated by tight-binding models, which shows better agreement with the diffusive case. Figs. \ref{fig:expkwant}(b,c) plots the same data as in Fig. 1(d) in the main text, fitted by the ballistic model Eq.\ref{eqn:IcnTballis} and the diffusive model Eq.\ref{eqn:Icdubos}, respectively. Similar to the tight-binding calculation, $E_{Th}$ and $T_{c,n}$ have similar values.

\begin{figure}[h!]
  \centering
    \includegraphics[width= 0.5\textwidth]{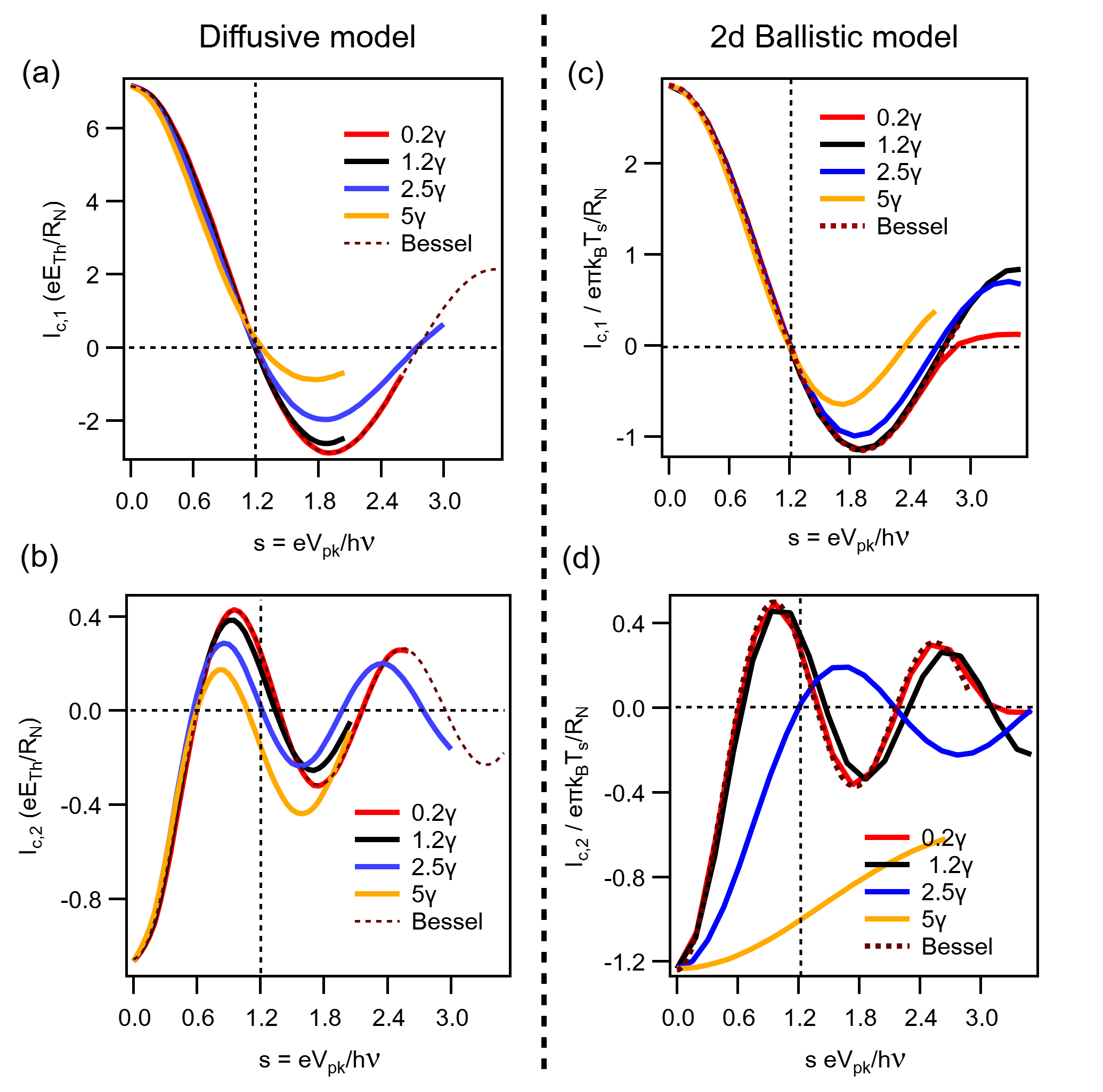}
\caption{Comparison between ballistic and diffusive models under irradiation: (a, b) The calculated $I_{c,1}(s)$ and $I_{c,2}(s)$ from Usadel equations with finite relaxation rate $\gamma = 1.2E_{Th}$.  $T = 0.005E_{Th}/k_B$. $\Delta = 50 E_{Th}$. Same as Figs. 2(e,f) in the main text. (c,d) $I_{c,1}(s)$ and $I_{c,2}(s)$ calculated by quasiclassical Green's function in the ballistic limit with finite relaxation rate $\gamma = 0.4k_B T_s/h$. $T = 0.002T_s$. [$L = 0.2$, $\hbar v_F$ = 1, $\Delta$ = 10. Thus $\xi = \hbar v_F/\Delta = 0.1$ and $k_B T_s = \hbar v_F/2\pi (L+\xi)$ = 0.53.] Bessel functions $J(s)$ and $J(2s)$ are ploted for (a, c) and (b, d) respectively (dashed lines)}
\label{fig:Usadelballis}
\end{figure}

The difference between diffusive and ballistic regimes is more clearly differentiated in the supercurrent response to irradiation, calculated by the quasiclassical Green's function approach \cite{greens,PauliMicrow}. Figs. \ref{fig:Usadelballis}(a,b) reproduce the Figs. 2(e,f) in the main text by the time-dependent Usadel equations incorporating finite relaxation rate. $I_{c,1}$ and $I_{c,2}$ are the Fourier coefficients of the CPR, and $s = eV_{pk}/h\nu$  is the normalized irradiation power ($V_{pk}$ is the peak voltage of the irradiation). Similarly, the time-dependent Eilenberger equations with finite relaxation can be constructed for the ballistic system \cite{EschrigMicrow}. The results are plotted in Fig. \ref{fig:Usadelballis}(c,d). As shown by the tight-binding simulations, the characteristic temperature $k_B T_s = \hbar v_F/2\pi (L+\xi)$ plays a similar role in the ballistic regime to that of $E_{Th}$ in the diffusive regime. The same irradiation frequencies as in Figs. \ref{fig:Usadelballis}(a,b) are used. The calculation conditions listed in the caption are normalized against $k_B T_s$. In order to be compared with the diffusive case, the $s$-axis for $I_{c,1}$ is rescaled so that all curves have zeros at $s$ = 1.2. The same scaling factor is then used for each $I_{c,2}$ of respective irradiation frequency. The ballistic results show qualitative difference to diffusive results and the experiments. The calculation assumes perfectly ballistic system and perfect contact transmission. In reality, small disorder and imperfect contact may lead to $I_{c,1}(s)$ and $I_{c,2}(s)$ more similar to the diffusive case.

\section{Susceptibility calculation using Kubo formalism and tight-binding model}

From the Kubo formalism, the total $\chi$ is \cite{Trivedi}:

\begin{equation}
\begin{aligned}
& \chi = \chi_J + \chi_D + \chi_{ND} \\
& \chi_J = \frac{2\pi}{\Phi_0}\frac{\partial I_s}{\partial \varphi} \\
& \chi_D = -\frac{i\nu_r/\gamma}{1-i\nu_r/\gamma}\displaystyle\sum_n \left(\frac{\partial\epsilon_n}{\partial\varphi}\right)^2\frac{\partial f_n}{\partial\epsilon_n} \\
& \chi_{ND} = -\displaystyle\sum_{n, m \neq n} |J_{nm}|^2\frac{f_n-f_m}{\epsilon_n-\epsilon_m}\frac{ih\nu_r}{i(\epsilon_n-\epsilon_m-h\nu_r)+h\gamma}
\label{eqn:chi}
\end{aligned}
\end{equation}
\noindent where the three terms are called ``Josephson term", ``diagonal term", and ``non-diagonal term" respectively. $\chi_J$ is purely in-phase with the ac flux modulation and is proportional to the phase derivative of the supercurrent $I_s$. $\chi_{D}$ has both real and imaginary part. The finite imaginary part $\chi''_D = \operatorname{Im}\chi_D$ indicates that there is a out-of-phase response thus a dissipation. This diagonal dissipation is caused by relaxation of thermally excited ABSs via inelastic scattering and has distinct phase dependence in which $\chi''_D$ is 0 at $\varphi = 0$ and $\pi$ and has two peaks symetric to $\varphi = \pi$ \cite{BastienPRB}. This contribution is important at high temperature ($k_BT\sim E_g$) and is maximum when the frequency of the ac flux $\nu_r$ is comparable to the inelastic scattering rate $\gamma$. In the experiment $k_BT\ll E_g$ and $\nu_r\ll \gamma$, this term is negligible. Indeed in the measured $\chi''(\Phi)$ without irradiation, no phase dependence expected from $\chi''_D$ is seen. The third term $\chi_{ND}$, resulting from inter-level transitions induced by microwave photons, also has real and imaginary part, and $\chi''_{ND} = \operatorname{Im}\chi_{ND}$ corresponds to dissipation. $\chi''_{ND}$ is the same as defined in the main text and is the dominant contribution to the total $\chi''$. We note that total dissipationless $\chi'$ which is directly proportional to the measured resonance frequency shift $\delta \nu_r$ is equal to $\chi_J+\chi'_{ND}$, where $\chi'_{ND}$ is the real part of $\chi_{ND}$. Defining $\delta\chi = \chi(\pi)-\chi(0)$, in the limit $k_BT\sim h\nu_r\ll h\gamma \lesssim E_{Th}$, it is shown that $\delta\chi'_{ND} \lesssim \delta\chi''_{ND} \sim (h\nu_r/\gamma)\delta\chi' \ll\delta\chi'$ \cite{BastienPRB} and thus the total $\chi'$ can be approximated by $\chi_J$. From the main text, indeed we see $\delta\chi' \approx 20\delta\chi''$, justifying the integration of the total $\chi'(\varphi)$ to obtain $I_s(\varphi)$. In summary, in the experimental condition, the measured susceptibility can be approximated as $\chi' \approx \chi_J$ and $\chi'' \approx \chi''_{ND}$.

For the tight-binding calculation of $\chi''_{ND}$, the system is constructed using a square lattice with $W\times L = 50\times30$ sites and strong disorder $D = 1.5\Delta$. In order to include all terms with significant $(f_n-f_m)/(\epsilon_n-\epsilon_m)$ while improving computational efficiency, we cut off the energy at $\sim 0.8\Delta$ (larger than the microwave irradiation energy $h\nu = 0.5\Delta$). The matrix element of the current operator $J$ is computed by \cite{meydi}:
\begin{equation}
\begin{aligned}
J_{nm} & = \langle n|(\hbar/i)\nabla|m\rangle \\
& = \frac{\hbar}{i}\displaystyle\sum_{j,j'} \langle n|x_j,y_j\rangle\langle x_j,y_j|\nabla|x_{j'},y_{j'}\rangle\langle x_{j'},y_{j'}|m\rangle \\
& =  \frac{\hbar}{i}\displaystyle\sum_{j} \Psi^{e*}_n(x_j,y_j)[\Psi^{e}_m(x_j+1,y_j)-\Psi^{e}_m(x_j,y_j)] \\ 
&  +\Psi^{h*}_n(x_j,y_j)[\Psi^{h}_m(x_j+1,y_j)-\Psi^{h}_m(x_j,y_j)]
\label{eqn:chitight}
\end{aligned}
\end{equation}

\noindent where $\Psi^{e}_n(x_j,y_j)$ and $\Psi^{h}_n(x_j,y_j)$ correspond respectively to the electron and hole components of the wavefunction at site $j$.

\section{Low power irradiation data}

\begin{figure}[h]
  \centering
    \includegraphics[width= 0.5\textwidth]{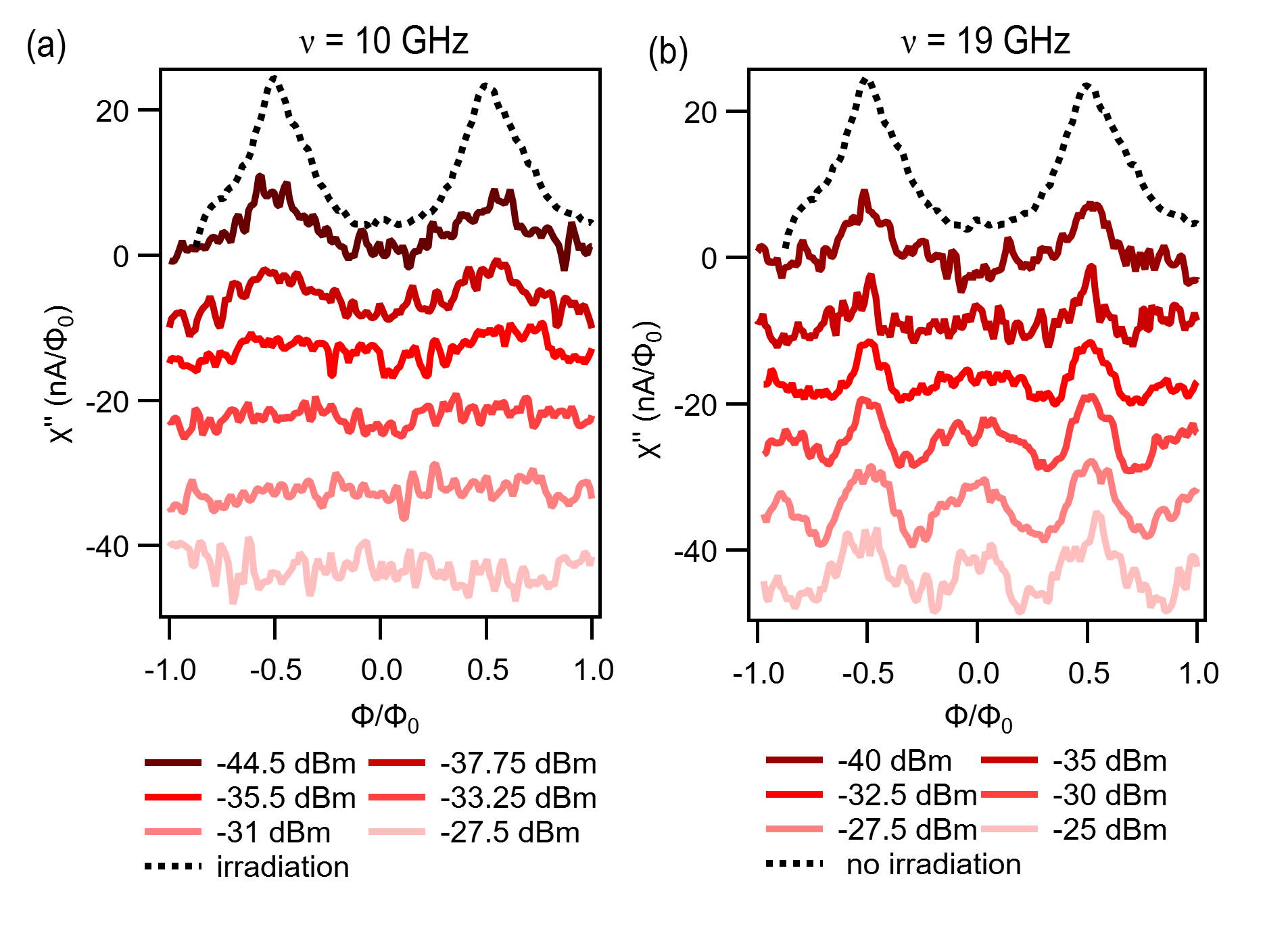}
\caption{$\chi''(\Phi)$ at low irradiation power: (a) $\nu$ = 10 GHz. (b) $\nu$ = 19 GHz. Data is taken at $T$ = 12 mK and $V_g$ = 1 V.}
\label{fig:lowP}
\end{figure}

In the main text, the lowest irradiation power for both $\nu$ = 2 GHz and 19 GHz flattens $\chi''(\Phi)$. Fig. \ref{fig:lowP} displays another dataset taken at $T$ = 12 mK and $V_g$ = 1 V starting from smaller irradiation power. Fig. \ref{fig:lowP}(a) is taken at $\nu$ = 10 GHz, and there is a gradual decrease in $\chi''(\Phi_0/2) - \chi''(0)$ as $P$ increases and at high power $\chi''(\Phi)$. Fig. \ref{fig:lowP}(b) is taken at $\nu$ = 19 GHz. There is also a gradual decrease in $\chi''(\Phi_0/2) - \chi''(0)$ as $P$ increases, while at high power $\chi''(\Phi)$ exhibits the halved periodic features same as the main text Fig. 3(c).

